\documentclass[epjST]{svjour}
\usepackage{times, epsfig}
\usepackage{amssymb}
\usepackage{amsmath}
\usepackage{graphicx}
\usepackage[latin1]{inputenc}
\usepackage{graphicx}
\usepackage{setspace}
\usepackage{psfrag}
\usepackage{cite}
\usepackage{multirow}
\usepackage{wasysym}
\usepackage{multirow}
\usepackage{float}
\usepackage{pdfsync}
\usepackage{pstricks}
\usepackage{soul}				
\begin{document}

\title{Models for the modern power grid}
%
%
\author{Pedro H. J. Nardelli\inst{1}\fnmsep\thanks{\email{nardelli@ee.oulu.fi}} \and
Nicolas Rubido \inst{2,3} \and 
Chengwei Wang \inst{2} \and 
Murilo S. Baptista \inst{2} \and 
Carlos Pomalaza-Raez\inst{4} \and 
Paulo Cardieri\inst{5} \and 
Matti Latva-aho\inst{1}}
%
%
\institute{Department of Communications Engineering, University of Oulu, Finland \and 
Institute for Complex Systems and Mathematical Biology, SUPA, University of Aberdeen, UK \and 
Instituto de Física, Facultad de Ciencias, Universidad de la República, Uruguay \and
Department of Engineering, Indiana University -- Purdue University Fort Wayne, U.S. \and 
School of Electrical and Computer Engineering, University of Campinas, Brazil} 
%
%
\abstract{
%
{ This article reviews different kinds of models for the electric power grid
that can be used to understand the modern power system, the smart grid.
From the physical network to abstract energy markets, we identify in the literature
different aspects that co-determine the spatio-temporal multilayer dynamics of power
system.}
%
We start our review by showing how the generation, transmission and distribution 
characteristics of the traditional power grids are already subject to { complex
behaviour appearing as a result of the the interplay between dynamics of the nodes
and topology}, namely synchronisation and cascade effects.
When dealing with smart grids, the system complexity increases
even more: on top of the physical network of power lines and controllable sources of
electricity, the modernisation brings information networks, renewable intermittent
generation, market liberalisation, prosumers, among other aspects.
In this case, we forecast a dynamical co-evolution of the smart grid and other kind
of networked systems that cannot be understood isolated.
This review compiles recent results that model electric power grids as complex
systems, going beyond pure technological aspects.
From this perspective, we then indicate possible ways to incorporate the diverse
co-evolving systems into the smart grid model using, for example, network theory
and multi-agent simulation.
} 
\maketitle

%
\section{Introduction}
\label{sec_intro}
Electric power grid names the system deployed to deliver electricity from the
generating units (power plants) to { the} end users (houses, industries
etc.) \cite{Fang12}.
Such a system is traditionally a one-way network, where the generators ``produce''
the electric energy, that needs to ``travel'' long distances in high-voltage
transmission lines, to arrive in the distribution network
that delivers the electricity to the consumers.
These three blocks -- generation, transmission, and distribution -- have
their own engineering challenges (many of them already solved in both research and
implementation levels).

In any case, the power grid elements are interdependent, which \textit{per se} may
allow for complex processes such as cascade effects \cite{Hines11}.
Big blackouts exemplify them, where few small events occurring 
{ at approximately} the same time may lead to a collapse in the
whole interconnected system. 

The scenario described above illustrate the ``old-fashion'' electric power grids.
In contrast, modern grids -- the so-called smart { grids} \cite{Fang12} -- uses information and communication technologies (ICTs) to { collect information (such
as the behaviours of suppliers, consumers, and prosumers) and take actions in an
automated fashion to improve the efficiency, reliability, economics, and sustainability
of the production and distribution of electricity \cite{Appel12,USA2012}. The
smart-grid, still an idea on the designing table, would enhance the reliability of
the energy infrastructure, decrease the impact of supply disruptions and even
create an internet of energy. A change of paradigm in the way energy is produced, distributed, traded, and consumed, is set to take place.}

Technological development of signal processing and two-way communication techniques
{ will be} fundamental to securely distribute information among the grid elements.
Besides, few big and many small generators are expected to coexist in smart grids. 
This allows for more distributed energy sources (e.g. solar panels or wind turbines)
and self-sufficient micro-grids.
The control of these variable sources of energy is also a big technological challenge,
where real-time action is required in many situations \cite{Fang12}.

From another perspective those aspects of smart grids enable new products, services, and markets.
For instance, any person who has a solar panel at home may become an energy trader.
Therefore a new market appears within smart grids, where consumers,
or groups of consumers, play an active role in the electric power system: they have
evolved from simple consumers to active \textit{prosumers}. { The social collective
behaviour of the consumers and traders might bring drastic effects into the stability
of the grid.}

From this brief description, we can also foresee possible issues beyond technicalities.
For instance, markets are known to be unstable where few speculative agents (persons
or computer programs) may collapse the whole economic system -- we are nowadays
experiencing this in the current crisis triggered by the US housing market crash
\cite{Rooy12}. { This would not be different in the smart grid.}
For instance, one might think about speculative agents that buy energy when the price
is low, store it and then resell when the price is high.
If this is the case, how this simple speculative behaviour will change the power flows
inside the grid?
How the abstract market network affects, and is affected by, the physical power network?
The answers to those questions are unknown and generally even not investigated.
%

In this review paper, we attempt to describe a wide range of existing network models
for the structures of  smart grids beyond technical aspects and { discuss} how they
are inter-related, { aiming at bridging the gap between these models, each coming
from different disciplines.}   

To do this, we divide our presentation into different sections.
%
Section \ref{sec_phy} focuses on { physical models of the grid} and the { complex
behaviours} and events that { can} happen in the traditional grid, { such as
frequency synchronisation and cascade effects.}
Section \ref{sec_InfoNets} deals with the challenges of building a dedicated
communication network for smart grids.
In Section \ref{sec_Renew}, we present the benefit and { problematic issues} in
using renewable sources of energy and other emerging technologies that may change
the grid dynamics (e.g. electric vehicles).
Section \ref{sec_Contextual} reviews the complexity related to the smart grid
management and how it is context-dependent (e.g. the smart grids designed for rural
zones are different from the one{ s} for urban areas).
In Section \ref{sec_GeneralModels}, we { present theoretical approaches that can
be used to combine all these aspects of the energy system in order to model and study
the smart grid dynamics as a complex multiplex networks \cite{jesus}, a network of
networks formed by different layers, with different dynamical behaviours, and different connections.}
Section \ref{sec_Final} concludes this review, pointing out some possible ways to go.

%
\section{The physics of electric power grid}
\label{sec_phy}
The power grid networks are still subject to a balance between supply and demand,
or \textit{load balance}, since batteries and other related technologies that exist
nowadays can only store small amounts of electric energy.
As we have mentioned, power grids are topologically interconnected.
This fact, summed with the inherent need for load balance within the network,
characterizes complex dynamics of the grid.
In this section, we review some of the main results in the field. Specifically,  focusing on synchronization issues and cascade effects.

\subsection{Synchronization}
\label{subsec_sync}
In power grids, electricity is normally transmitted in alternating current (in 50 or
60 Hz depending on the country). {Therefore, in addition to
load balance, synchronisation between all elements is a crucial aspect to the system} \cite{Blaabjerg06}.
A famous description of how synchronisation can appear in networks of coupled
oscillators was made by Kuramoto in \cite{Kuramoto1975}.
Such description has been used as a model for many complex systems, ranging from
biological clocks to power grids.
For instance in \cite{Filatrella12}, the authors have shown that, if machines can be modelled as having the same dissipative coefficient $\alpha$, the same
moment of inertia, and power is not lost during transmission, then a power
grid where power-conservation laws holds can be cast into a second order Kuramoto-like
model. Namely,
\begin{equation}
\label{Kuramoto_PowerGrid}
\begin{aligned}
	&\dot{\theta}_i = \omega_i,\\\vspace{2ex}
	&\dot{\omega}_i = - \alpha_i \; \omega_i + P_i + P_\textup{max} \sum\limits_{j=1}^{N} a_{ij} \sin\left(\theta_j-\theta_i\right),
\end{aligned}
\end{equation}
where $i=1,2,...,N$ represents a consumer or a generating unit of the system,
$\theta_i$ the time-dependent phase { of unit} $i$, $\omega_i$ is
the time-dependent frequency centred in the power grid one of $\Omega$ Hz, $P_i$ is
{ related to} the power consumed (if $P_i<0$) or generated (if $P_i>0$), $a_{ij}$
indicates whether two nodes are connected and $P_\textup{max}$ is related to the maximum
power that can be transmitted in a direct link, { $\alpha_i$ represents the damping coefficient of the machine $i$.  Since Eq. (\ref{Kuramoto_PowerGrid})
considers that the {{power dissipated}} in transmission is zero, {{i.e., whatever is transmitted is received}}, this model should be considered
as a model not for the whole grid, but as a model for the high-power network, responsible
for the transmission of power.}

{ Noticing that the uncoupled dynamics of Eq. (\ref{Kuramoto_PowerGrid}) is linear
 ($\omega_i$ and -$\alpha \omega_i$) and has constant parameters ($P_i$), and that the
magnitude of the term representing the power being transferred $P_\textup{max}$ is constant
for every edge, then, the linear stability of the synchronous solution ($\dot{\theta}_i
=\dot{\theta}_j$ and $\dot{\omega_i}=0$) is independent of the parameters that make
the network heterogeneous ($P_i$). If a further assumption is made \cite{Carareto13},
i.e., the network has nodes that are highly connected (large degree) and $P_\textup{max}$ is
sufficiently strong, resulting in a network whose nodes have small phase difference,
then the stability of the synchronous nodes is determined by the eigenvalues of the
matrix describing the orthogonalised variational equations: 
\begin{equation}
\left[ \begin{array}{cc}
0 & 1 \\
P_\textup{max} \gamma_i & -\alpha  
\end{array}
\right]
\label{matrix_stability}
\end{equation}
\noindent
where $\gamma_i$ are the eigenvalues of the Laplacian matrix of the network, defined
as $L_{ij}=A_{ij}-\mathbf{I}k_i$ ($A_{ij}$ is adjacency matrix, $\mathbf{I}$ is identity
matrix, and $k_i$ is the degree of node $i$). $\alpha$ is the average damping coefficient by assuming $\alpha_i=\alpha_j$, $\forall i,j$. In this sense, the Master Stability
Function \cite{Pecora98} is therefore given by the values of  $P_\textup{max}$ and $\alpha$
such that the matrix (\ref{matrix_stability}) has only eigenvalues with negative real
part. This function allows one to understand how the topology of the grid naturally
(without control) favour synchronous states. 
{{Notice that 
Eq. (\ref{matrix_stability}) is obtained by expanding $\cos$ about zero (due to the small phase difference), which means that 
the power-grid is operating in a mode where small amounts of power are transferred in the cable lines.}} 
In  \cite{Carareto13}, the authors
have also studied long-term global stability by numerical means, by determining coupling
strengths for which the long-term behaviour of Eq. (\ref{Kuramoto_PowerGrid}) is
frequency synchronisation, even after large perturbations are introduced in the system.}
Their results were applied to study the natural stability of the Brazilian power grid.
Using the same grid model, the authors in \cite{Rohden12} studied the collective
dynamics in large-scale oscillatory networks, using the British grid to
exemplify their results.
Interestingly the authors showed a somehow counter-intuitive result: the more
decentralized the power sources are, the more robust the grid is
against topological faults. 
On the other hand, they also showed that a more decentralised system is at the same
time more sensitive to perturbations.

In \cite{Witthaut12} Witthaut and Timme have shown how the inclusion of new links can affect the power grid synchronisation.
The authors argument is that new links may lead to Braess's paradoxes \cite{Braess05}
-- { phenomenon} where adding new links \textit{reduces} the overall performance
of the system.
The authors recently extended this work in  \cite{Witthaut13} where they characterise
other consequences of such a paradox in the network such as cascade effects (which is
the topic of the next subsection).

{ The swing equation was used in \cite{Motter13,Dorfler12} as the dynamic model of generators: $\frac{2H_i}{\Omega}\ddot{\theta}_i=P_{mi}-P_{ei}-\frac{D_i}{\Omega}\omega$. The parameters in the swing equation have a clear physical meaning. $H_i$ is the inertia constant of the generator, $P_{mi}$ is the mechanical power provided by the generator. $P_{ei}$ is the electrical power transmitted to the network from the generator reading $P_{ei}=E_i'\sum_{j=1}^NE_j'(B_{ij}\sin(\theta_i-\theta_j)+G_{ij}\cos(\theta_i-\theta_j))$, where $E_i'$ is the internal voltage of $i^{th}$ generator, $B_{ij}$ and $G_{ij}$ are the conductance and susceptance between node $i$ and $j$. $D_i$ is the damping coefficient with typical values between 0 and 2. In the Kuramoto-like model, the parameters are normalised. For example, $P_i$ is proportional to the power produced or generated, but not equal. If the power dissipation on the transmission line is set to zero (i.e. $G_{ij}=0$), power grids modelled by the swing equation become algebraic equivalent to power grids modelled by the Kuramoto-like equation ( \cite{Filatrella12}). One only needs to divide the constant value $\frac{2H_i}{\Omega}$ on both sides of the swing equation, noticing that $\dot{\theta_i}=\omega_i$. There is however a fundamental difference between both models regarding the way consumers are modelled. Consumers are described in the Kuramoto-like model ( \cite{Filatrella12}) by simply setting $P_i<0$. If one wants to model the dynamics of a power grid by the swing equation, one models a consumer by a unit that uses either a constant impedance, power, or current. A reduced network is obtained by eliminating load buses (responsible to bring power to the consumers). This reduced network has values of $B_{ij}$ and $G_{ij}$ that incorporate the existence of consumers.}

{ 
We cite the following recent works that deal with synchronization of power grids as a
{ complex dynamical network}: \cite{Motter13,Dorfler12,Dorfler13,Porco13,Lozano12}.
In particular, in  \cite{Motter13}, the authors have derived a type of Master
Stability Function for the Swing equation, an equivalent mechanical model for a power
system, more general than the model in (\ref{Kuramoto_PowerGrid}), since it takes into
consideration that power is dissipated in the power lines. Therefore, it can be used to
study the stability of a power network that models not only the transmission but also
the distribution of energy. The authors have shown this type of Master Stability
Function  in terms of the eigenvalues of a matrix constructed from the eigenvalues of
the coupling matrix and parameters of the generators. The coupling matrix is a matrix
that contains terms related to the adjacency matrix (topology), but also terms due to
the phase difference between the nodes (dynamics). Even though the stability can then be
determined in terms of the network topology, the information obtained from one type of
network cannot be extended directly to another arbitrary network topology. For this
reason, we coin this result as a  Master Stability { type}-Function, which in fact
can help one to understand how to enhance the stability of the synchronous state of a
particular network, by changing parameters but not the structure.  Another work of
relevance is the one in  \cite{Dorfler12}, which has studied the global long term
stability of the swing equation, neglecting the dissipation in the transmission of power.
Their result is general, but still is a function of the phase diference, and therefore
cannot be used to infer the stability of a power grid by studying another power grid.
Inferring the stability of a network out of another network is a crucial point if one
wants to grow a power network, an issue that becomes even more needed now that the grid
will need to be adapted to become the smart grid. Finally, in the work of 
\cite{Dorfler13}, the authors provide a general global long-term stability condition
for a mechanical system analogous to a structure preserving model for the power-grid.
This model considers that nodes have different dynamics. Each dynamical behaviour can
be seen as a model for the dynamics of different elements of a power-grid (see the
discussions of the work of  \cite{giraldo}, in Sec. \ref{sec_GeneralModels}) . Some
nodes are modelled by a second-order equation (similar to Eq, (\ref{Kuramoto_PowerGrid}))
and others are modelled by a first-order equation (similar to making $\dot{\omega}_i=0$,
in Eq, (\ref{Kuramoto_PowerGrid})). The condition states that the network will be
frequency synchronous $\omega_i=\omega_j$ or phase coherent ($|\theta_i-\theta_j|<\pi/2$)
if the eigenvalues of $L^+{\bf \omega}$ are smaller or equal then $\sin{(\gamma)}$, where
$L^+$ is the pseudo-inverse of the Laplacian matrix of the network, and 
$|\theta_i-\theta_j|\leq\gamma<\pi/2$.}

\subsection{Cascade effects}
\label{subsec_cascade}
Cascade effects denotes a series of sequential events that is triggered by one or
few initial disturbances within the system, causing in some cases its collapse
\cite{Osorio09}.
In power grids, many blackouts are known to be caused as a result of cascades of
small outages events \cite{Dobson09,Hines09,Vaiman12,Pahwa14}.
%
The best reported examples are the blackouts in North America in 2003, Europe in 2006,
Brazil 2009 \cite{Vaiman12}, { and India 2012 \cite{zhang2013}}, where a sequence
of outages shut down large parts of the electric grid.
%

Within complex systems, cascade effects are closely related to its topology
\cite{Strogatz01,Watts02,barabasi2000,pagani2013}: random, preferential-attachment,
small-world, or { scale free lead to particular cascade characteristics.
For example, scale-free  networks are known to be unresilient
against selective attacks. In particular, scale-free networks are more sensitive to
attacks on short-range than on long-range links \cite{lai2004}.

While there is still no consensus about whether there is a typical topology for real
power grids, in recent studies it was found that the power grid follows an exponential
distribution of degrees \cite{Albert04,pagani2013}. There are however exceptions. In
 \cite{Chassin05} the authors have shown evidences that the North America power grid
could have a scale-free topology \cite{Chassin05}. Contributing to this controversy, the
authors in   \cite{Sanchez12} compared the data from the US power grid using
different fitting models. Based on the topological distance analysis, they have shown
that the power grid is neither scale-free nor small-word. In fact, it is shown that
there are topological differences in the grids worldwide, as demonstrated by the works
in \cite{Witthaut13,Crucitti04,Casals07}. It is worth to mention that
\cite{pagani2011,pagani2013}, the authors have suggested that the controversy about the
type of network seems to exist in the topology of the high tension power network, while
the medium and low voltage networks seem {\it far} from being a small-world network.

The analyses carried out in  \cite{hines2010} are based on the ``electric distance'',
the equivalent resistant of a Kirchoff network. The authors in this work claimed 
that they are able to better capture the specificities of power grids (where Ohm and
Kirchoff laws prevails) than when topological distances or connectivity measures are
employed.  The electric distance, whose analytical calculation was derived in
\cite{Rubido13a}, indeed provides a powerful way to understand the flows and cascade
effects in networks.}
The electrical distance between two nodes -- also known as equivalent resistance --
multiplied by the amount of flow entering a network is the load capacity of the edge,
where load and flow are linearly related.
It is also shown in \cite{Rubido13b} that  one strategy to avoid cascade effects in DC
models of power grid is to design a network topology such that its input flows { avoid surpassing} the individual load capacities of the
edges.

%
Using the electric distance, Eppstein et al. proposed in \cite{Eppstein12} an algorithm
to identify events that cause cascading failures in a power { system}.
This work was further extended in \cite{Sanchez13} for partitioning the power grid
networks that may be used as a tool for avoiding cascade effects.
In a recent paper \cite{Quattrociocchi14}, the authors proposed a self-healing strategy for Kirchoff networks, which can be potentially employed to avoid cascade fails in power grids.

In most of the cases, cascade failures are related to connections { or nodes that
are dropped in the power system}.
However, as described before, power grids are also subject to Braess paradoxes when the
topology of the network changes \cite{Witthaut13}.
%
The authors showed that the \textit{inclusion} of a link may also lead to cascade events
and illustrate this in the British grid (Fig. 6 in \cite{Witthaut13}).
At this point, we have illustrated that the traditional power grids face a complex
dynamics even when looking only at simplified systems.
%

{ In most of the works presented in the previous section, it is was usual practice to
neglect the dissipation in power lines (the dissipative term $\alpha$ appearing in Eq.
(\ref{Kuramoto_PowerGrid}) refers to the dissipation in the machines).  Additional
analytical difficulties appear if one wants to take into account power loss due to
dissipation. The work in  \cite{Motter13} has considered dissipation in the power
lines. However, only local linear stability of the frequency synchronisation is stated
in this work. By neglecting dissipation, theoretical results obtained from models of
the power-grid such as the ones from  \cite{Carareto13,Witthaut12,Dorfler13} remain
only valid to study the power transmitted throughout the high-voltage transmission lines.

Cascade models provide a way understand how the resilience of a grid is related to both
its topology and the dissipation in its power lines. In  \cite{haipeng}, the authors
have studied a model of a power-grid whose power lines have a distribution of heterogeneous effectiveness (power-loss). The authors showed results of loads
redistributions after failures of nodes in the UK power grid.
They showed that if generators fail in a power network with the same topology of the UK
power grid, the cascading effect appearing would be less abrupt than the one in a network
with a scale-free topology. As mentioned, scale-free networks are unresilient to selective attacks, such as an attack only
on the generators. Compared with some existing models, the proposed model in 
\cite{haipeng} requires larger node capacity to suppress cascading failure, being more
realistic to actual power grids.}

\section{Information networks and smart grids}
\label{sec_InfoNets}
The information and communication technologies (ICTs) are changing the dynamics of our society \cite{Helbing12a}.
For example, the internet, which was built to connect researchers back in 1980, is now part of everyday life \cite{Leiner09}.
Together with its benefits, the internet is also creating new issues related to security, privacy, criminal activities amongst others \cite{Hoven12}. 
New generations tend to be globally networked, which produces new challenges, new risks and even changes in the scales of events \cite{Helbing13a}.
And yet, new concepts that machines, appliances, humans etc. must be always connected have been developing building the internet of things \cite{Atzori10} or even the internet of everything \cite{Becker13}.

In this way, humans and its social relations start being even more relevant to understand the system behaviour \cite{Giannotti12}. 
The complexity involved in ICT-based systems is expect to be much higher than in ``traditional'' ones; for instance, tendencies tend to propagate faster as people are able to exchange more information and then time-scales of events tend to decreases, which in turn may facilitate cascade effects  \cite{Helbing13a}.
These topics have been recently covered in a whole special issue of the \textit{European Physical Journal Special Topics} \cite{Bishop12}.

These kind of ICT-based solutions are also in the core of the smart grid -- some people even claim that smart grid will become the internet of energy \cite{Appelrath12,Marsan12}.
The authors in \cite{Xu13} analysed wireless mesh networks for time-critical communications in smart grids, targeting always the minimal delay.
In \cite{Galli11} Galli et al. provided an extensive analysis of power line communication in the context of smart grids, also indicating the importance of the grid topology since ``(...)the power grid is not only the information source but also the information delivery system''.

As a reference in the topic, we cite \cite{Gungor13} as a complete survey on the technological challenges to design a communication network suitable for smart grids.
They described systematically (refer to Fig. 2 therein) the advantages and drawbacks of possible technologies, from different wireless systems to power line communication.
%
Interestingly, they indicated the communication requirements needed for different smart grid applications (e.g. demand response, outage management, distribution automation, advanced metering etc.) as well as the best communication standards for each context. 
%


From another perspective, the authors in  \cite{Yan13} stated the challenges in the deployment of communication systems for smart grids: complexity, efficiency, reliability and security.
In the following sections, we will cover some of these challenges and try to understand how  the electric grid and the communication networks may co-evolve in different contexts. 

\section{Challenges of renewable energy and other technologies}
\label{sec_Renew}
%
The discussions about CO$_2$ emissions and their effects in { climate change} have been growing \cite{Jacobson09}.
One of the key points to decrease such emissions regards the use of renewable sources of energy instead of oil- or coal-based sources.
Looking to the consumption side, electric/hybrid vehicles are seen as another important change to decrease emissions.
In this section, we will review the challenges in implementing these technologies\footnote{A counterpoint of those solutions are found in the analysis provided by Zahner and compiled in \cite{Zehner12}, where the author claims that such celebrated solutions are only ``green illusions''.}.

\subsection{Renewable energy}
\label{subsec_Renew}
In \cite{Jacobson09} Jacobson identified the following options as the most important renewable sources of electricity:
``solar-photovoltaics (PV), concentrated solar power, wind, geothermal, hydroelectric, wave, tidal, nuclear, and coal with carbon capture and storage technology''.
Herein we focus on the characteristics of PV and wind sources due to their wider use nowadays.

PV technology is used to convert solar radiation into electricity using semiconductors that present photovoltaic effect \cite{Parida11}.
While PVs have been employed for some time in space industry as the source of energy to spacecraft, its commercial use start growing with the energy crises in the 70's \cite{Nelson03}.
%

{ Wind power plants} is the technology that uses the wind movements to produce energy.
As a source of electricity, it dates 1888 \cite{Carlin03}, where ``(...)the Brush wind turbine in Cleveland, Ohio had produced 12 kW of direct current power for battery charging at variable speed''.

In both cases, the idea is to capture the natural potential of energy and convert to electricity so that the dependency of ``dirty'' sources like oil and coal decreases.
In this way, the objective is to switch the large-scale oil/coal power plants to ``clean'' PVs and wind-mills technology.
However, as a { natural phenomenon}, wind and sun shine are unpredictable in small-scales and consequently their integration in the power grid network may cause instability in the power supply (remind the need for load balance within the power grid): sometimes peaks of production will not be followed by the demand, or vice-versa  \cite{Carrasco06}.

Synchronization issues described in the previous section would also worsen when such renewable sources integrate in the grid. { This can be clearly seem if we study the global mean field of Eq. (\ref{Kuramoto_PowerGrid}). Defining the global mean field as 
$\mathbf{Y}=<\theta_i> = \frac{1}{N}\sum_{j=1}^N\theta_j$, 
and noticing that $\frac{1}{N}{\sum_{i=1}} \sum_{j=1}^Na_{ij}\sin{(\theta_j-\theta_i)} = 0$,  then the mean field equation for the Kuramoto-like power-grid 
network is given by 
$\ddot{\mathbf{Y}} = -\alpha \dot{\mathbf{Y}} + \langle P_i \rangle$. If $\langle P_i \rangle=0$, meaning that demand matches production of energy, the mean field equation describes a steady-state solution, and the system is stable. If however $\langle P_i \rangle \neq 0$, for example 
$\langle P_i \rangle >> 0$ due to a peak of generation, then the mean field variable becomes unstable and might oscillate. Each node in the 
grid is somehow coupled to the global mean field. That causes oscillations in $\omega_i$ making nodes to deviate from the natural frequency of the grid $\Omega$.  To see this, we define the local mean field of node $i$ as $\langle \theta_i \rangle=\frac{1}{k_i}\sum_ja_{ij}\theta_j$, and the coordinate transformation $r_ie^{i \langle \theta_i \rangle} = \frac{1}{k_i}\sum_ja_{ij}e^{i\theta_j}$. {Neglecting imaginary terms \cite{restrepo2005}}}, equation (\ref{Kuramoto_PowerGrid}) can be written in terms of the local mean field as 
\begin{equation}
\ddot{\theta}_i = -\alpha \dot{\theta}_i + P_i + P_\textup{max}k_i r_i \sin{(\langle \theta_i \rangle - \theta_i)}
\label{transformed_kuramoto}
\end{equation}
\noindent
where $\langle \theta_i \rangle = \mathbf{Y} + \xi_i(t)$. Therefore, oscillations in $\mathbf{Y}$ might destabilise the machine $i$, making its frequency to largely deviate from the natural frequency of the grid.

These engineering aspects are in fact the target of many research groups worldwide.
For example, the authors in \cite{Xiangjun13} propose a battery-based approach to smooth the supply variations in PV and wind generators.
In \cite{Riffonneau11} Riffonneau et al. studied a dynamical programming approach to deal with power flow when PV generators are present in the power grid.

All these solutions have in common the advance of storage technologies to filter the unbalances inherent of PVs and windmills \cite{Bevrani10}.
In other words, the problem is that these generators have a dynamic that depends on factors out of engineering control such as presence of clouds in the region where the PV is installed or wind speed in the region of the wind-mill.
Therefore, the engineering problem is to build a network that can cope with these aspects while the power grid maintain its desirable characteristics (e.g. load balance, synchronisation etc.) .
As we saw in Section \ref{sec_phy}, these aspects are closely related to the network topology  and its power flows.
Later in Section \ref{sec_GeneralModels} we will come back to this topic when providing the guidelines to assess the system dynamics.

\subsection{Electric vehicles}
\label{subsec_ElectircVehicles}
Another important element in the modern grids is the electric vehicle (EV) and its unique load characteristics \cite{Wu11}: EVs can be seen as a mobile battery that can be plugged anywhere in the power grid to be charged or even provide electricity in demand peaks.
In other words, the topology of the network may be highly mobile if EVs start being widely used.
This indicates that the challenges regarding load balance and power flows will become even more complex.

One possible solution is to use scheduling algorithms to make use of the benefits of EVs while avoiding charging them in demand peaks \cite{Wu12}.
Hilshey et al. proposed in \cite{Hilshey13} an algorithm to cope with the impact of charging EVs based on the so-called smart charging methods.
Marshall et al. { gave} in \cite{Marshall13} another interesting perspective of the problem by showing the effects of EVs on the heat transfer  in the electric cables.
In \cite{Nyns10}, the authors { assessed} the impact of charging vehicles in residential regions, comparing different ways to cope with it.
Spatio-temporal dynamics of EVs were analysed in \cite{Bae12} based on deterministic fluid equations, queueing systems and stochastic models.

Some other studies { evaluated} the impact of EVs based on actual data.
For example, the authors prepared in \cite{Meyer07} a report about how EVs affect the regional US power grid.
In some chapters of the book \cite{MacKay08}, MacKay { dealt} with many technical aspects of EVs based on vehicle specification or available power grid data.
As pointed in \cite{MacKay08}, EVs are not only limited to cars but also collective transportation networks like trams, trains etc.
In this case, one should understand much more than technical aspects related to EVs and further evaluate their impact looking at traffic patterns of people and goods in the transportation network, which is closely related to when and where the vehicles need to be plugged in the power grid.
In other words, the dynamics of the transportation network will affect, and will be affected by, the power grid.

\section{Contextual aspects of smart grids}
\label{sec_Contextual}
In recent years there has been a considerable amount of work carried out in several countries to understand the best ways to modernize their power grids \cite{US09,Giordano11,CGEE12}. 
These studies have mainly focused on the technical aspects of the proposed smart grid, showing several real case studies that have been used to evaluate its expected advantages.
Most of them, however, { only evaluate} the deployment of smart meters across a neighbourhood or a small city, using then the information collected from those meters to manage the energy use, in cooperation with the consumers. 

The inclusion of consumers -- who are people part of social contexts -- is indeed a key aspect in the success of smart grids.
In fact, many results about network sciences have been first presented by social scientists \cite{Wasserman94}.
For example, the famous six-degrees of separation that characterizes small-world topology were introduced empirically by Milgram, a social psychologist, in \cite{Milgram67}.
After this seminal work, many other models of social networks have been proposed, as indicated in the Chapter 3 of \cite{Newman09}.

Processes of diffusion in networks are also very important to understand how a technology such as the smart grid applications will propagate or not throughout the social system.
Diffusion have been extensively studied in epidemiology to understand how diseases spread over a network based on social relations (refer to Chapter 17 of \cite{Newman09}).
Dealing with new technologies, Rogers studied the diffusion of innovations in the society \cite{Rogers10}.

Market (de)regulation is an important point in the deployment of smart grids.
Many specialists claim that real-time pricing or other market-based incentive strategies can indeed change behaviours of consumption, decreasing the demand for electricity in peak periods \cite{Rad10,Conejo10,Saad12,Qian13}.
In other words, the smart grid system would be designed upon market strategies to flat the demand load curves.
Consumers who only demand electricity in the traditional electricity market would be also expected to be able to trade (buy and sell) energy.
In this case, the new \textit{prosumers} can use the technologies described in Section \ref{sec_Renew} to produce energy and sell the surpluses to the grid. 

While these solutions are optimal under the optics of the mainstream (neoclassical) economic theory, they may fail when implemented in actual systems \cite{Helbing13b}.
Amongst the main critiques, we can cite rationality and equilibrium as assumptions that produce interesting mathematical results \cite{Bompard2011}, but many times fail to model the actual socio-economical { phenomenon}. 
Some authors go even further: the utility theory of value -- which is the basis of neoclassical economic theory -- cannot be either proved or disproved since their basic assumptions cannot be tested \cite{Cockshott09}.
Without going into the merits of these works, we can conjecture some possible issues in deregulated electricity markets such as speculation (e.g. buy electricity when it is cheaper, use batteries to store it and sell when it is more expensive) or rational, autonomous demand control (e.g. turn off air-conditioning when it is too hot or the heating when it is too cold). { These issues might contribute to stabilise the market, but might also destabilise it. The energy market can present the same problematic issues that stock markets have. For example, phenomena such as the "Minsky moment". The longer the period of stability, the larger the potential risk for greater instabilities when market participants must change their behavior.  This effect in the stock market could cause a major collapse in the value of energy.}

Yet from the demand side, the inner city dynamics as well as its long-term evolution are important to understand the grid implementation.
When researching the ``science of cities'', social science researchers, together with urban planners and physicists, have been developing analytical and simulation tools to study the evolution of urban areas as in \cite{Batty07,Batty13,Bettencourt13}.
Their focus has been on large cities trying to find universal features for places across the world.
The main conclusion from those studies is that factors such as land use, transportation, and communications play very important roles in the social dynamics of cities, which in turn affects the energy consumption and the power grid dynamics \cite{Carreno11,Melo12}.

The different results presented in this section suggest that the success of the smart grid goes beyond the already complex technological challenges.
In this case, there is an increasing awareness that the proper study, modelling, and planning of smart grids should not be limited to the technical aspects.
In this way, the research should include not only the power grid network itself, but also the different networks that affect, and are affected by,  the power system. 
These studies should also include the social, economical and other contextual networks where  the smart grid is operating as pointed out in \cite{Carreno11,Melo12,Ribeiro12,Brummitt13,Nardelli13,Vries12}.  
All in all the generation, transmission and distribution of electricity do not exist on its own isolated domain. 

\section{Modelling smart grids}
\label{sec_GeneralModels}
We have described many challenges related to the modernisation of the electric power grids.
This section indicates possible ways to include these diverse factors into an unified framework inspired by cutting-edge research in network theory, multi-agent simulations and other tools developed to assess complex systems. 
Following this line of thought, researchers from different fields participated in the workshop ``Power Grids as Complex Networks'' (Santa Fe Institute on May, 2012).
The conclusion of such a transdisciplinary event was \cite{Brummitt13}: ``Integrating and transcending disciplines will enable new ideas to shape the power system(...)''.

For example, Kremers presented in his doctoral thesis \cite{Kremers13} a way to model smart grids as agent-based simulation, including diverse aspects such as people behaviour using home appliances and time series equations to model the wind-power generators in a given place.
A research group from the US Department of Energy at Pacific Northwest National Laboratory have been developing the \textit{GridLAB-D} agent-based simulator for distribution power system \cite{Chassin08}.
The authors in \cite{Divenyi13} recently studied the control of renewable sources of energy using agent-based simulation.

In \cite{Carreno11}, the authors proposed a cellular automaton to model the expansion of a city so as to predict the spatial demand of electricity based on the theory presented in \cite{Batty07}.
This work was further extended in  \cite{Melo12}, modelling the city expansion using multi-agent models instead of the simple automaton.
In both cases, however, the modernisation aspects of the power grid (as the ones cited in Sections \ref{sec_InfoNets} and \ref{sec_Renew}) have been neglected, and the authors only explore the growing in consumption based on the city growth.

{ In  \cite{pagani_PHYSICAA2014}, the authors study how different
network topologies and growth models contribute to a network appropriated to a decentralised electricity market. 
They showed which is an optimal growth evolution strategy that balances best the ratio
between increased connectivity and costs.}

Other possibility to model smart grids comes from the last advances in control theory for networked systems, regarding the 
observability of a system, i.e., the partial or full determination of the state variables of a system from limited measurements. 
For instance, the authors in \cite{Yang12} have analysed the observability of power grids based on the so-called ``network observability transition'', indicating when observable islands appear in the system as the number of measurement nodes, randomly selected, increases. These islands are group of nodes that can have their states known (complex voltages) by a reduced number of observations using phasor measurement units.

In \cite{Brummitt12} Brummitt et al. studied how interdependence between networks can suppress cascade effects in power grids.
They showed based on the Bak-Tang-Wiesenfeld model \cite{Bak88} that there exist a number of interconnections that minimize the risk of cascade effect whereas the systems still benefit of being interconnected.

Along these lines, a recent work \cite{Ribalta13} has proposed a tensorial mathematical formalism to tackle multiplex networks where nodes that are connected in many different ways. 
They have extended the quantities characterizing traditional one layer networks such as degree centrality, modularity, clustering coefficient, eigenvector centrality, to multilayer networks, where each layer represent a different type of connections between nodes.
Other results on network theory applied in smart grids were presented back in Section \ref{sec_phy}, when discussing synchronization and cascade effects.

{ Finally, the work in  \cite{giraldo} offers a bridge into how engineering and physics can come together to understand the power-grid with a communication structure. They 
showed how a power-grid composed of generators (modelled by the swing equation), nodes representing the AC energy obtained from DC generation (coming from wind and solar), and load buses (that distributes the power) can be controlled by  applying linear feed-back controllers in a reduced number of nodes. By control, it is meant that one can either achieve frequency synchronisation (as previously defined, $\dot{\theta}_i=\dot{\theta}_j$)
or phase coherence ($|\theta_i-\theta_j| < \gamma$). The authors showed how these states can be reached as a function of the control strength and eigenvalues of a Laplacian matrix constructed from the power-grid structure and the structure of the network formed by the nodes that are being observed and controlled. They also provided conditions for the control when the observations are sampled, a situation that can often be present when for example communication is interrupted.}

\section{Final remarks}
\label{sec_Final}

\begin{figure}[t]
\centering
\includegraphics[width=0.75\columnwidth]{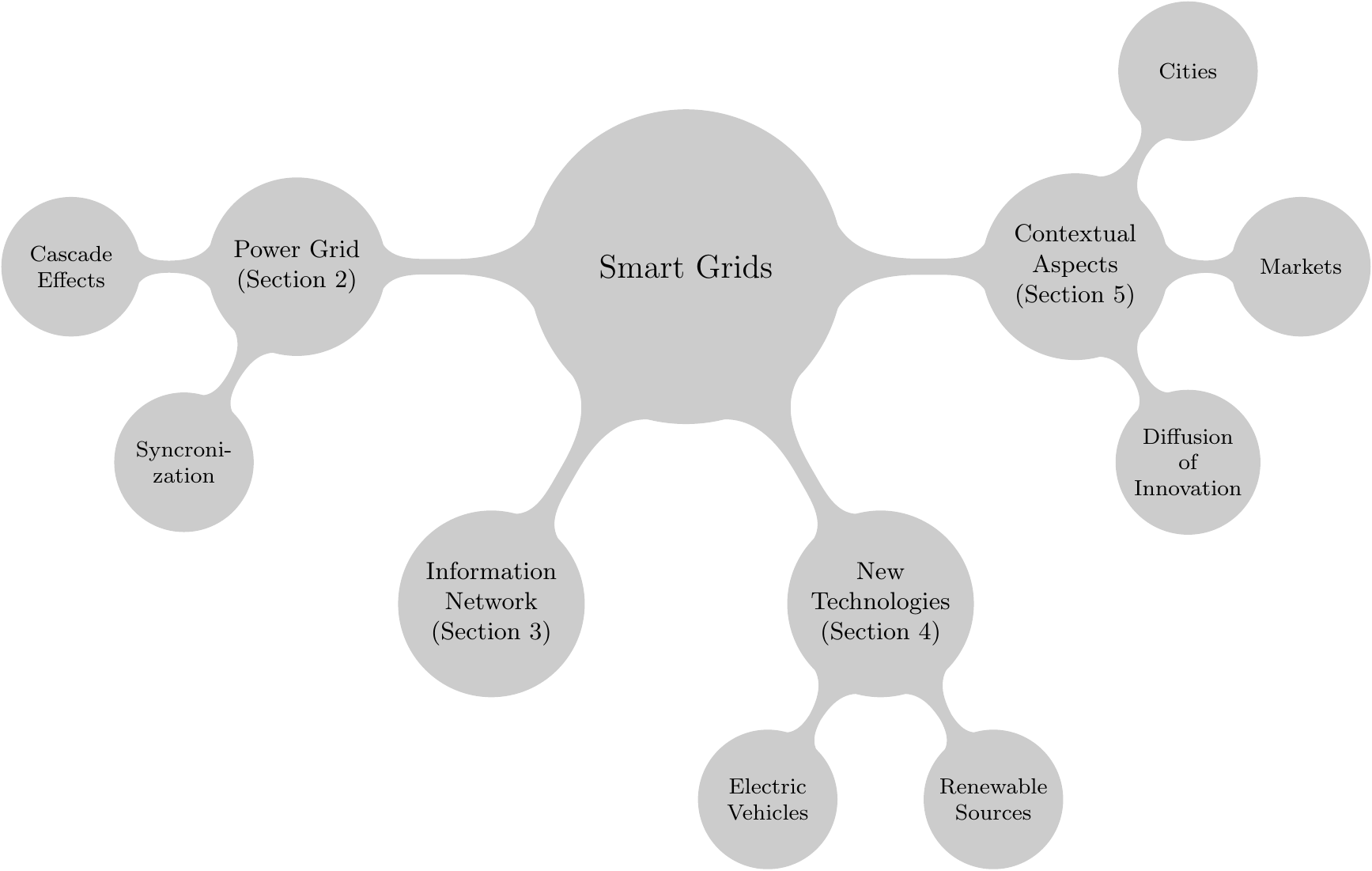} 
\caption{Schematic figure to represent some aspects of the smart grid that can be modelled as (complex) networks. Each topic is related to the respective section it appears in this review.}
\label{fig_mindmap}
\end{figure}

In this review article, we presented many challenges that arise with the modernization of power grids. 
From purely technical issues of the topology of the generation, transmission and distribution networks (Section \ref{sec_phy}) to social issues of  electricity market deregulation, we revisited different phenomena related to smart grids that affect each other in their co-evolving dynamics.
We also identified in the recent literature of networks and multi-agent systems the tools needed to tackle these challenges.
Figure \ref{fig_mindmap} compiles the topics covered herein.

To conclude this paper, it is worth saying that the understanding of the smart grids is still limited -- although growing -- due to the lack of transdisciplinary knowledge in the field.
%
We expect that this review also serves as a roadmap of possible ways to carry out research in smart grids. { In  Sec. 1-5, we present the approaches being used to model sub-units or elements that will be part of the smart grid. In Section \ref{sec_GeneralModels}, we present theoretical approaches and ideas of how these sub-units and elements could be theoretical merged into a unifying framework in order to understand the behaviour of the power system as a complex multiplex networks \cite{jesus}, a network of networks formed by different layers, with different dynamical behaviours, and different connections. A complex network that is smart: resilient, energy efficient, stable, flexible and cheap to evolve, making the energy to arrive at the end-user with a reduced amount of greenhouse production.}  

\acknowledgement{This work was partly supported by the Science without Boarders Special Visiting Researcher fellowship CAPES/Brazil 076/2012, SUSTAIN Finnish Academy and CNPq/Brazil 490235/2012-3 jointly funded project, CNPq/Brazil 312146/2012-4. MSB acknowledges EPSRC grant EP/I032606/1.}

\end{document}